# A network-constrained rolling transactive energy model for EV aggregators participating in balancing market

Peng Hou, Guangya Yang, Junjie Hu, Philip J. Douglass

*Abstract*—**The increasing adoption of renewable energy sources increases the need for balancing power. The security concerns of the distribution system operators are increasing due to fast adoption of distributed energy resources (DERs). So far, various operational models are proposed to manage the DERs for economic benefit. However, there is lack of an efficient operational model that balances the security, market, and uncertainty management. With the focus on the aggregator's operation, this paper developed an operational model for continuous operation of aggregators interactively with the grid operators. A rolling optimization model is developed to manage the uncertainties from prediction errors. The network security is ensured continuously through an iterative price negotiation process between the distribution network operator and the aggregators. The optimality of the solution is guaranteed by convexification of the mixed-integer formulation.**

| | |
|---|---|
| ***Acronym*** | |
| UR/DR | up-regulation/down-regulation respectively |
| ***Parameter*** | |
| $i, j, k, r, g$ | Time slot index, aggregator index, EV number index, rolling optimization process index, and bus no index, respectively |
| $\mu^{DA}$, $\mu^{BM}$ (Dkk/kWh) | Predicted DAM and BM price |
| $\eta_{Ch}$, $\eta_{Dis}$ (%) | Charging/discharging efficiency |
| $T_k^{st}$, $T_k^{st}$ (h) | The start and end charging time of the $k^{th}$ EV |
| $N_{EV}$, $N_{RBM}$, $N_{RDA}$ | Total number of EVs, rolling window, and length of rolling window |
| $E_b$ (kWh) | Capacity of battery |
| $SOC_{init}$, $SOC^{min}$, $SOC^{max}$, $SOC_{des}$ (%) | Initial, minimum, maximum SOC, and desired SOC, respectively |
| $P^{MaxCh}$, $P^{MaxDis}$ (kW) | Maximum charging and discharging power of EV, respectively |
| $c_d$ (Dkk) | Degradation costs |
| $c_{bat}$ (Dkk) | EV battery capital cost |
| $L_{ET}$ | battery life in terms of energy throughput (kWh) |
| $L_c$, $DoD$ | Cyclic life and depth-of-discharge respectively |
| $m^{up}/m^{down}$ | Upper boundaries of UR and DR respectively |
| $N_{bus}$ | Total number of buses of the system |
| $M_g$ | Participation factor representing the participation preference of each EV owner. |
| $U^{max}$, $U^{min}$ | Minimum and maximum voltage limits |
| $U^0$ | Initial voltage of the buses of the network |
| $P_{trans}^{Max}$ | Capacity of transformer |
| $J_{21}^{-1}$ | Submatrix of the inverse Jacobian. |
| $\beta$ | Step length, set as 0.4 initially |
| $\omega$ | Convergence standard, set as 0.005 |
| ***Variables*** | |
| $\delta^a$, $\delta^b$ | Binary variables representing charging and discharging modes respectively |
| $P^{Ch}$, $P^{Dis}$ (kW) | Charging and Discharging schedules of an EV, respectively |
| $\delta^1$ , $\delta^2$ | Binary variables flagging the UR and DR operation |
| $P^{up}$, $P^{down}$ (kW) | Powers for UR and DR operation, respectively. Both are positive values always. |
| $\delta^3$ | The battery is in charging mode. $\delta^3\ (P^{Ch}-P^{Dis}-\delta^1\ P^{up}+\delta^2\ P^{down})\geq 0$ |
| $\delta^4$ | The battery is in discharging mode. $\delta^4\ (P^{Ch}-P^{Dis}-\delta^1\ P^{up}+\delta^2\ P^{down})\leq 0$ |
| $z^1=\delta^1\ \delta^3\ P^{up}$ (kW) | An EV provides UR and in charging mode. |
| $z^2=\delta^2\ \delta^3\ P^{down}$ (kW) | An EV provides DR and in charging mode. |
| $z^3=\delta^1\ \delta^4\ P^{up}$ (kW) | An EV provides UR and in discharging mode. |
| $z^4=\delta^2\ \delta^4\ P^{down}$ (kW) | An EV provides DR and in discharging mode. |
| $P^{FCh}$, $P^{FDis}$ (kW) | The optimal charging and discharging schedules of an EV in DAM respectively |
| $P^{BM}$ (kW) | The optimal schedule of the aggregator for participating BM |
| $P^{ChN}$, $P^{DisN}$ (kW) | New schedules of EV after TE deviating from $P^{BM}$ due to NC |
| $P^{DSO}$ (kW) | The optimal schedule of DSO in TE market |
| $\lambda$ (dkk/kWh) | Virtual price signal |
| $P_{g,i,j}^{ChN*}$, $P_{g,i,j}^{DisN*}$ (kW) | New schedules after TE deviating from $P^{BM}$ due to NC of bus g associated with aggregator j at time interval i |
| $P_{g,i,j}^{DSO*}$ (kW) | The optimal schedule of DSO after TE at bus $g$ associated with aggregator $j$ at time interval $i$ |

## I. INTRODUCTION

The Paris agreement underlines the urgency of decarbonizing the current energy sector [1]. With the integration of distributed energy resources (DERs) such as solar photovoltaics (PV), batteries, and electric vehicles (EVs), the distribution system operation becomes more complex and dynamic. Congestion and voltage constraints are the main security challenges to the system operators to ensure a safe operation, while the flexibility in the new type of demand provides the system operator possibilities to address these challenges [2]. Small prosumers usually lack knowledge, information, and resources to optimize their assets, which naturally make their operation best conducted through a representative party such as aggregator. The EV aggregator (EVA) is actually an intermediary who manages the energy consumption of total subscribed EV owners.

Efforts have been given in vehicle charging, demonstrating various algorithms to optimize vehicle charging strategy considering uncertainties in prices and energy demand [3][4], or network constraints (NCs) [5][10]. In [6], the optimal charging strategies for EVs from either the perspective of DSO or commercial parties were proposed. A close-loop EV charging strategy for loading serving entity was proposed in [7] which can adjust the real-time prices in response to EV consumers' behavior. To maximize the EVs' profits in both day-ahead market (DAM) and reserve markets, a linear programming (LP) model was studied in [8]. With the increasing number of EVs, the DAM price may be influenced by the EV charging demand which formulated as an aggregative game model in [9].

To make the best use of EVs, optimal scheduling of EVA was investigated in [11]-[13] where the EVAs acted as a balancing responsible party (BRP). In [11], the optimal operating strategy of EVA participating in the DAM, intraday market (IM), and BM was proposed considering the stochastic behavior on vehicle parking patterns and battery lifecycle costs. Authors in [12] proposed an optimal schedule for EV taking the battery discharging degradation costs, the network constraints and different EV usage behaviors into account. A co-optimization approach for both customer and system operator's objectives was presented in [13].

It should be noted that the above energy scheduling are all based on the perspective of one party with a centralized optimization problem. In this model, the NC were managed either by the aggregator or the distribution system operator (DSO). For example, a distribution locational marginal price was obtained centrally and broadcasted by the DSO to alleviate the congestion problem in [5]. This type of model is hard to guarantee the customers' willingness and available resources in response, as well as the fairness of the price due to the special network structure in distribution grids.

On the uncertainty management, to reduce the errors' impacts on decision making, robust optimization has been adopted in scheduling, eg [16]. Robust optimization only leads to conservative actions but not the uncertainly level. Rolling window optimization (RWO) was introduced in the literature to utilize the latest information for reduced uncertainty [17]. The EVs charging schedule in DAM considering its impact on the unit commitment schedule was formulated as a mixed-integer linear programming (MILP) in [18] through a day-by-day rolling analysis. Taking the advantages of RWO and model predictive control, a real-time microgrid dispatch for a combined cooling, heating, and power was presented in [19].

Considering the above-mentioned work, the paper developed further the network constrained transactive energy (NCTE) concept [14][15][20] by taking the advantage of RWO in uncertainty handling for real-time operation and convexification of the aggregators' model to ensure the optimality. The framework enables the participation of EVAs in Nordic DAM and balancing market (BM) by optimal schedules of EVAs considering the latest demand status and distribution system constraints. The technical improvement in comparison to the previous work is fourfold: 1) RWO method is applied for optimizing the operational strategy of each aggregator to reduce the impacts of electricity price prediction errors; 2) The battery operating cost is modeled and considered in the optimization problem formulation; 3) Taking the V2G into account, a new bi-linear optimization model for EVA is proposed which can be applied both for DAM as well as BM; 4) A convexification method is developed to ensure the convexity of the formulation.

The organization of this paper is as follows. In Section II, the Nordic market structure and the RWO framework of EVA are introduced. In Section III, the mathematical modeling of the aggregators' optimal operational strategy in the DAM and BM are presented at first, then the procedure for obtaining the TE solution using distributed optimization is specified. The simulation results and discussions are given in Section IV. Section V presents the conclusions.

## II. Market Participation Procedure For EVA

In this section, the structure of the Nordic electricity market is introduced at first. Thereafter, the application of RWO in each market is specified. The relation among each player in TE and the overall optimization framework of the aggregator is given in the end. In this work, the problem formulation does not include network utilization tariffs. Including these would affect the profitability of V2G.

### A. *Nordic Market Structure*

In Northern Europe, the power system has the main five distinct markets: financial market, DAM, IM, BM, and ancillary services market [21]. The first three markets are operated by Nord Pool while the TSOs maintain the system balance through the last two markets [22]. DAM and BM are pool while IM is bilateral market. Because historical price data from bilateral contracts are not publicly available, IM modelling is usually not done. This study focuses on the operation of the aggregator in the DAM and BM.

### B. *EVA operation in the market*

We consider EVA will first operate in DAM. In DAM, EVA will estimate the entire day energy and power requirement of the EVs under subscription, the past customs, and forecasted electricity price to schedule the energy requirement. This energy requirement is rather primitive with varying accuracy according to the actual operating time situation, which certainly will not be close to be optimal. No matter what optimization technique it is used for DAM participation, the deviation on the electricity price, EV availability, and the expected energy requirement will occur in various degrees which need be corrected when the time is close to the actual operating hours by different parties depending on the mechanisms and incentives. There are different possibilities in the current market frame for the energy providers to correct their own imbalance for better economy before the actual operating hours, eg IM, or spot market. In this work, we consider EVAs will sell their imbalance as a service to the grid operator.

### C. *RWO design for Balancing Market*

Due to the large integration of renewable energy, as the hour of operation approaches, the BM's price is likely to deviate from the DAM price. The BM begins when the day starts. EVA is starting the planning of BM participation in the hour before the real operating hour starts. The RWO time window is 3 hours here, but it can be stretched. The proposed RWO method for aggregators participating in BM is illustrated in Fig. 1.

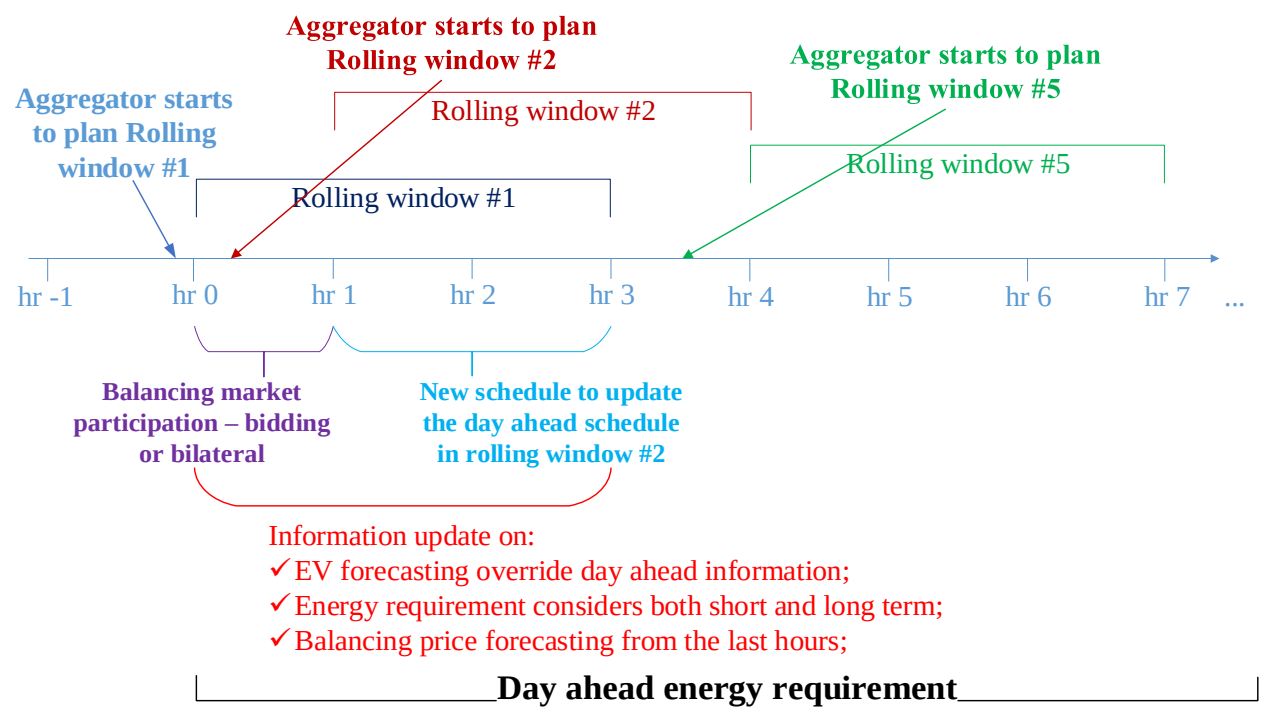

Fig. 1. Rolling procedure for BM participation.

In this framework, EVA will plan for the next three hours BM participation given the updated information from the EV availability, updated state of charge (SOC), charging requirements (eg charger type) and regulating power price. It is assumed that we have a better prediction of the price in BM for the first coming hour that the last two hours, therefore, only the first hour decision is taken for BM participation. In the next hour, EVA will start the same process for the coming time window planning. The process will continue which makes it rolling-window-based operation.

The RWO takes advantage of the fact that the prediction error for BM prices is reduced closer to the hour of operation.

## D. *Transactive energy framework*

The principle of TE is shown in Fig. 2, which is similar to [20]. DSO and EVA is interacting through an independent market operator (IMO) for exchanging a virtual price signal. The total EVs can be viewed as a time-varying virtual storage (TVVS) model.

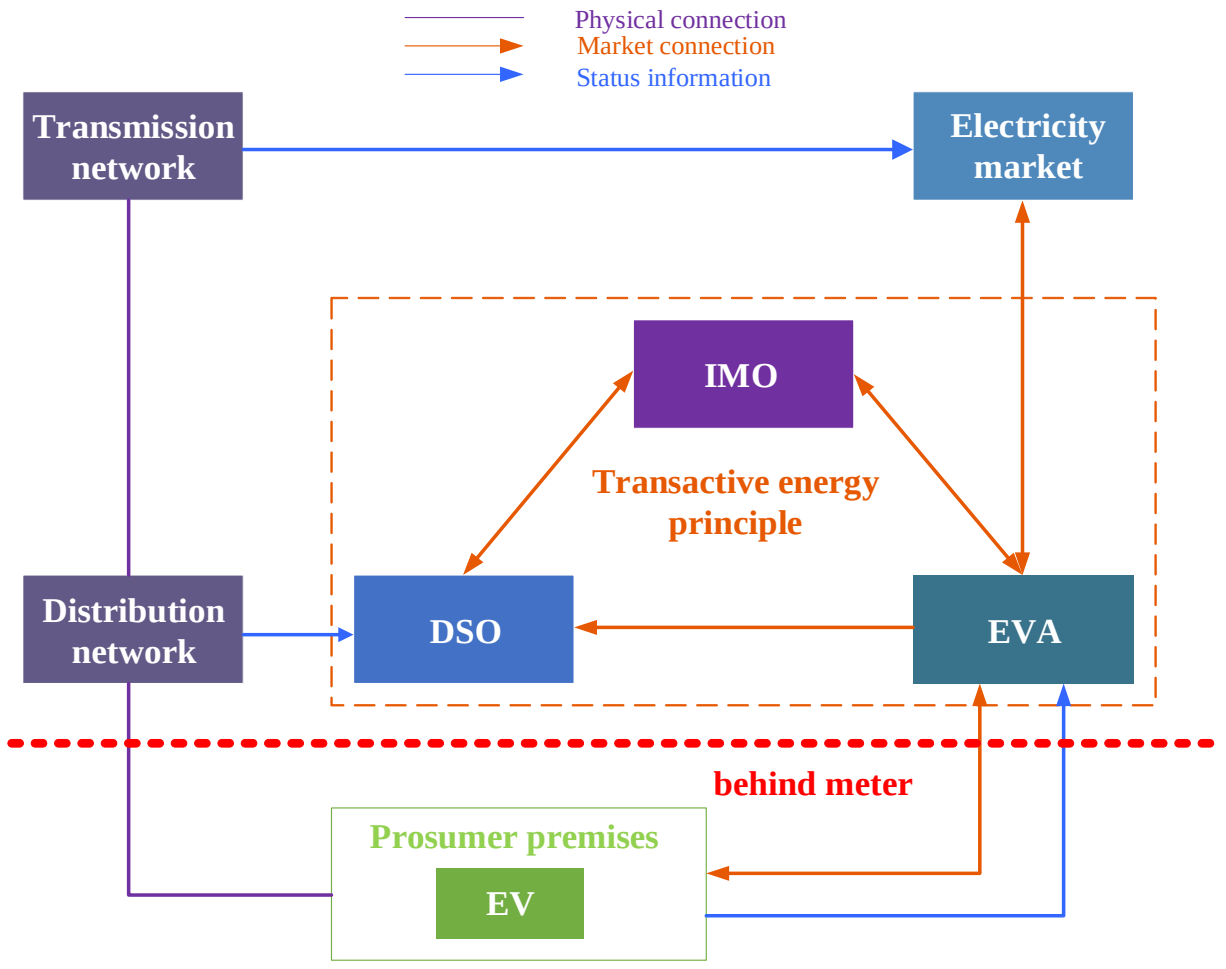

Fig. 2. Electricity Market (DAM, BM) of EVA in the distribution network.

## E. *Overall Optimization Framework for EVA*

Considering the constraints from the DSO, the optimization framework for EVA should be modified to incorporate the TE. The overall flowchart is shown in Fig. 3. In this study, assumptions are adopted as follows: 1) EVAs can receive information of energy needs, prediction of the market prices for the DAM and BM participation. This can be guaranteed with the current market transparency and the registered information of the public and private chargers; 2) The up and down regulation power supplied by EVA will be taken by TSO and fully activated by the TSO within one hour; 3) Distribution line capacity is not considered, and the power factor of the load is 1. This is verified by analyzing over 2000 smart meter data over 1.5 years in Denmark.

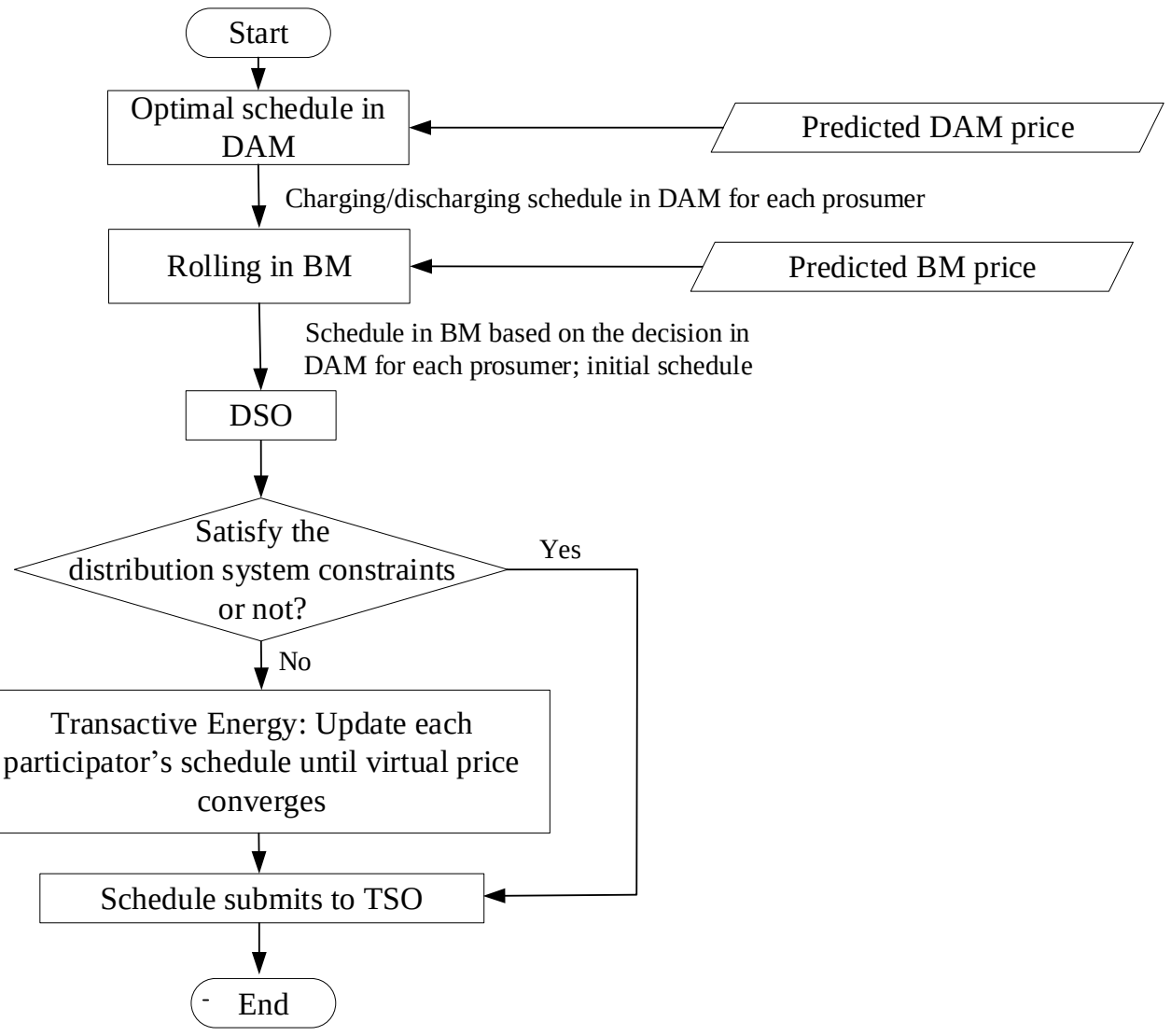

Fig. 3. The optimization framework for the aggregator.

## III. Mathematical Formulation

In this section, the mathematical formulation of the problem considering BOCs for EVAs in DAM and BM is first presented. Secondly, the aggregator's and DSO's objective when the network security constraints are violated are introduced. Then, the overall optimization problem is formulated as a social welfare maximization (and cost minimization) problem. The distributed optimization method is presented at last for decomposing the problem and solving it decentralized.

## A. *Optimization in DAM*

The EV prosumers are heterogeneous, having different driving patterns, charging/discharging efficiency, etc. In this work, the prosumers' behaviors are considered as 1) Different prosumers' daily driving route and working hours are represented by the different initial SOC, charging/discharging start and end time. 2) Since different prosumers may choose different types of EVs, the EV capacities and maximum charging/discharging power are modeled to be different.

Since the V2G function is expected to be active in DAM, two binary variables, $\delta^a_{k,i}$ and $\delta^b_{k,i}$, are introduced to indicate the EV operating charging and discharging status at a particular hour. Then, the proposed model is written in the following.

$$min \sum_{k=1}^{N_{EV}} \sum_{i=T_{st}(k)}^{T_{end}(k)} \left( P^{Ch}_{k,i} - P^{Dis}_{k,i} \right) \lambda^{DA}_{i} + c_{d,k} P^{Dis}_{k,i} \quad (1)$$

$$\text{S.t.} \qquad \delta^a_{k,i}, \delta^b_{k,i} \in \{0,1\} \quad (2)$$

$$\delta_{k,i}^{a} + \delta_{k,i}^{b} \le 1 \quad (3)$$

$$0 \le P_{k,i}^{Ch} \le \delta_{k,i}^{a} P_{k}^{MaxCh} \quad (4)$$

$$0 \le P_{k,i}^{Dis} \le \delta_{k,i}^{b} P_{k}^{MaxDis} \quad (5)$$

$$SOC_{k,i+1} = SOC_{k,i} + \left(P_{k,i}^{Ch}\eta_{Ch} - P_{k,i}^{Dis}{\eta_{Dis}}^{-1}\right)E_b^{-1} \quad (6)$$

$$SOC_{k,1} = SOC_{init}^{k} \quad (7)$$

$$SOC^{\min} \le SOC_{k,i} \le SOC^{\max} \quad (8)$$

$$SOC_{k,end} \ge SOC_{des} \quad (9)$$

The energy level of EV battery for each time slot is calculated by Eq. (6) and limited by Eq. (8). The expected SOC of each EV should be met at the end of the charging period is described as Eq. (9). The above problem is formulated as a MILP model and can be solved with the existing solver which will be introduced at the end of this section.

### *B. Battery operating cost (BOC)*

The battery lifetime is usually counted by the number of cycles and is highly related to the depth of discharge. In [12], the battery discharging degradation cost has been specified. In this work, the battery operating cost in addition to energy purchasing cost considering both the impacts of charging and discharging at an average DoD is formulated as follow:

$$c_d = c_{bat} L_{ET}^{-1} \quad (10)$$

$$L_{ET} = L_c E_b DoD \quad (11)$$

### *C. RWO in Balancing Power Market*

The up and down regulating power is decided in BM through RWO in this work considering the settled schedule in DAM. Some binary variables are defined to indicate the operational status of the battery and the type of regulation provided. It should be noticed that the above optimization problem contains bi-linear terms which make the problem non-convex. This intractable problem can be resolved by introducing extra artificial variables $z_{k,i,r}^{1}$ to $z_{k,i,r}^{4}$. This is the so-called big M method [23]. Through the big M linearization, the MILP formulation for EVA's optimal schedule in BM can be expressed in the following.

$$\min_{j \in \Omega_j} \sum_{r=1}^{N_{RBM}} \sum_{k=1}^{N_{EV}} \sum_{i=T_k^{st}}^{T_k^{end}} \Big[ \left(z_{k,i,r}^{2} - z_{k,i,r}^{1} + z_{k,i,r}^{4} - z_{k,i,r}^{3}\right) \lambda_{i+(r-1)N_{RL}}^{BM} + c_{d,k}\left(\delta_{k,i,r}^{4}\left(P_{k,i}^{FCh} - P_{k,i}^{FDis}\right) - z_{k,i,r}^{3} + z_{k,i,r}^{4}\right) \Big] \quad (2)$$

S.t.

$$\delta_{k,i,r}^{1}, \delta_{k,i,r}^{2}, \delta_{k,i,r}^{3}, \delta_{k,i,r}^{4} \in \{0,1\} \quad (3)$$

$$\delta_{k,i,r}^{1} + \delta_{k,i,r}^{2} \le 1 \quad (4)$$

$$\delta_{k,i,r}^{3} + \delta_{k,i,r}^{4} = 1 \quad (5)$$

$$0 \le P_{k,i,r}^{Up} \le P_{k}^{MaxCh} + P_{k,i}^{FCh} - P_{k,i}^{FDis} \quad (6)$$

$$0 \le P_{k,i,r}^{Down} \le P_{k}^{MaxCh} - P_{k,i}^{FCh} + P_{k,i}^{FDis} \quad (7)$$

$$0 \le \delta_{k,i,r}^{3}\left(P_{k,i}^{FCh} - P_{k,i}^{FDis}\right) - z_{k,i,r}^{1} + z_{k,i,r}^{2} \le P_{k}^{MaxCh} \quad (8)$$

$$-P_{k}^{MaxDis} \le \delta_{k,i,r}^{4}\left(P_{k,i}^{FCh} - P_{k,i}^{FDis}\right) - z_{k,i,r}^{3} + z_{k,i,r}^{4} \le 0 \quad (9)$$

$$SOC_{k,1,1} = SOC_{init}^{k} \quad (10)$$

$$SOC_{k,1,r} = SOC_{k,N_R+1,r-1}^{k} \quad r \in [2, N_{RDA}] \quad (11)$$

$$SOC_{k,i+1,r} = SOC_{k,i,r} + \Big[\left(\delta_{k,i,r}^{3}\left(P_{k,i}^{FCh} - P_{k,i}^{FDis}\right) - z_{k,i,r}^{1} + z_{k,i,r}^{2}\right)\eta_{Ch} + \left(\delta_{k,i,r}^{4}\left(P_{k,i}^{FCh} - P_{k,i}^{FDis}\right) - z_{k,i,r}^{3} + z_{k,i,r}^{4}\right){\eta_{Dis}}^{-1}\Big]E_b^{-1} \quad (12)$$

$$SOC_{k}^{\min} \le SOC_{k,i,r} \le SOC_{k}^{\max} \quad (13)$$

$$SOC_{k,end,r} = SOC_{des} \quad (14)$$

The constraints added to the formulation due to the utilization of big M linearization are written as follows [24]:

$$0 \le z_{k,i,r}^{1} \le p_{k,i,r}^{up} \quad (15)$$

$$m_{k}^{up} = P_{k}^{MaxCh} + P_{k,i}^{FCh} - P_{k,i}^{FDis} \quad (16)$$

$$-z_{k,i,r}^{1} - \left(\delta_{k,i,r}^{1} + \delta_{k,i,r}^{3}\right) \le -P_{k,i,r}^{up} + 2m_{k}^{up} \quad (17)$$

$$-z_{k,i,r}^{1} \le -P_{k,i,r}^{up} + m_{k}^{up} \quad (18)$$

$$z_{k,i,r}^{1} - m_{k}^{up}\delta_{k,i,r}^{1} \le 0 \quad (19)$$

$$z_{k,i,r}^{1} - m_{k}^{up}\delta_{k,i,r}^{3} \le 0 \quad (20)$$

$$0 \le z_{k,i,r}^{1} \le P_{k,i,r}^{down} \quad (21)$$

$$m_{k}^{down} = P_{k}^{MaxCh} - P_{k,i}^{FCh} + P_{k,i}^{FDis} \quad (22)$$

$$-z_{k,i,r}^{2} + m_{k}^{down}\left(\delta_{k,i,r}^{2} + \delta_{k,i,r}^{3}\right) \le -P_{k,i,r}^{down} + 2m_{k}^{down} \quad (23)$$

$$-z_{k,i,r}^{2} \le -P_{k,i,r}^{down} + m_{k}^{down} \quad (24)$$

$$z_{k,i,r}^{2} - m_{k}^{down}\delta_{k,i,r}^{2} \le 0 \quad (25)$$

$$z_{k,i,r}^{2} - m_{k}^{down}\delta_{k,i,r}^{3} \le 0 \quad (26)$$

Similar 'big M constraints' should also be added regarding $z_{k,i,r}^{3}$ and $z_{k,i,r}^{4}$ following the format in (25)-(36). Due to the paper length limitation, this part is ignored. It should be noticed that the schedule settled in DAM is a contract of purchasement instead of real-time delivery. Hence, it is possible to 'update' the EVs' schedules in BM which correspond to the new boundary described in (16) and (17). The final schedule of each EV after BM should be allocated to its corresponding bus so that the DSO's constraints can be considered. In this sense, the unconstrained schedule is defined as $P_{g,i}^{BM}$ which is the aggregated schedule of each bus *g* at time *i*.

### *D. EVA's Optimal Schedule*

The interest of EVA is to minimize the payment in the market. However, due to the system constraints' limitation, it has to update its own schedule to meet the DSO's requirement, which incurs the deviations from its optimal schedule. The willingness of EVAs to shift its schedule is incentivized by the virtual price signal released by IMO which will be explained in section III. F to G. The new optimization problem for an aggregator j is formulated as follows:

$$A_j = \min \sum_{g=1}^{N_{bus}} \sum_{i=\min(T_{st})}^{\max(T_{end})} \left(P_{g,i,j}^{ChN} - P_{g,i,j}^{DisN} - P_{g,i,j}^{BM}\right)^2 M_{g,i,j} \quad (27)$$

S.t.

$$0 \le P_{g,i,j}^{ChN} \le P_{g}^{MaxCh} \quad (28)$$

$$0 \le P_{g,i,j}^{DisN} \le P_{g}^{MaxDis} \quad (29)$$

$$SOC_{g,i+1,j} = SOC_{g,i,j} + \left(P_{g,i,j}^{ChN}\eta_{Ch} - P_{g,i,j}^{DisN}{\eta_{Dis}}^{-1}\right)E_b^{-1} \quad (30)$$

$$SOC_{g,1,j} = SOC_{init}^{g} \quad (31)$$

$$SOC_{g,end,j} = SOC_{des}^{g} \quad (32)$$

$$SOC_k^{\min} \le SOC_{k,i} \le SOC_k^{\max} \quad (33)$$

A large $M_{g,i,i}$ indicates a stronger willingness of participation in the BM. In this work, it is assumed that $M_{g,i,i}$ equals to 1 which means that for all the EV owners has the same willingness to participate in the BM. It should be noticed that the binary variables are neglected in the above formulation which is a convex optimization problem. The mathematical proof of convexification is given in the following.

**Proof:** Assume $P_t^+ \ge 0$ and $P_t^- \ge 0$ are the optimal charging and discharging solution satisfying $P_t^+ P_t^- \neq 0$, for any $t \in [1, N]$. $N$ is the total number of time slots. Let $Q_t^+ = P_t^+ - \varepsilon$ and $Q_t^- = P_t^- - \varepsilon\eta_+\eta_-$ to be another solution while $\varepsilon$ represents a small positive value. Those two solutions yield the same change of state of charge as:

$$Q_t^+\eta_+ - \frac{Q_t^-}{\eta_-} = P_t^+\eta_+ - \frac{P_t^-}{\eta_-} \quad (34)$$

Considering the quadratic objective function in (37), the optimality does not stand if the following condition is met:

$$\sum_{t=1}^{N}\left(Q_t^+ - Q_t^-\right)^2 < \sum_{t=1}^{N}\left(P_t^+ - P_t^-\right)^2 \quad (35)$$

$$\Rightarrow \sum_{t=1}^{N}\left(Q_t^+ - Q_t^- + P_t^+ - P_t^-\right)\left(Q_t^+ - P_t^+ - Q_t^- + P_t^-\right) < 0 \quad (36)$$

Taking $Q_t^+ = P_t^+ - \varepsilon$ and $Q_t^- = P_t^- - \varepsilon\eta_+\eta_-$ into (46), we can then obtain:

$$\sum_{t=1}^{N}\left[\left(2P_t^+ - 2P_t^-\right)\varepsilon\left(1-\eta_+\eta_-\right)\right] > \varepsilon^2\left(1-\eta_+\eta_-\right)^2 \quad (37)$$

In reality, the efficiencies are lower than 1, in other words, $\varepsilon(1-\eta_+\eta_-)>0$. So (47) can be rewritten as:

$$\sum_{t=1}^{N}\left(2P_t^+ - 2P_t^-\right) > \varepsilon\left(1-\eta_+\eta_-\right) > 0 \quad (38)$$

As described in (40)-(43), there is a desired SOC level at the end of the scheduling which can be rewritten as follows:

$$SOC_{end} = SOC_{initial} + \sum_{t=1}^{N}\left(P_t^+\eta_+ - P_t^-\eta_-^{-1}\right)E_b^{-1} \quad (39)$$

$$\Rightarrow \sum_{t=1}^{N}\left(P_t^+\eta_+\eta_- + P_t^-\right) = \left(SOC_{end} - SOC_{initial}\right)E_b\eta_- > 0 \quad (40)$$

$$\sum_{t=1}^{N}\left(P_t^+ - P_t^-\right) > \sum_{t=1}^{N}\left(P_t^+\eta_+\eta_- - P_t^-\right) > 0 \quad (41)$$

It can be seen that the assumption does not stand and the convexification of the original problem works.

### *E. Target of DSO*

The responsibility of DSO is to meet the energy demand of each aggregator while ensuring the overall operational schedule meets the distribution system constraints. DSO's optimization problem can be written as follows:

$$D = min\sum_{g=1}^{N_{bus}}\sum_{i=\min(T_{st})}^{\max(T_{end})}[P_{g,i}^{DSO} - \sum_{j=1}^{N_{agg}}P_{g,i,j}^{BM}]^2 \quad (42)$$

S.t.
$$-P_{trans}^{Max} \le P_{g,i}^{DSO} \le P_{trans}^{Max} \quad (43)$$

$$U_{g,i}^{\min} \le U_{g,i}^{0} - J_{21}^{-1}P_{g,i}^{DSO} \le U_{g,i}^{\max} \quad (44)$$

In (54), there is an approximation that assumes a constant power factor. This approximation method has been demonstrated to be an effective way of calculating the voltage deviations in [14][25] and thus applied in this work.

### *F. Social Welfare Maximization*

It can be seen from the previous explanation that both EVAs and DSO are desired to minimize their costs while meeting the system constraints. From a social welfare point of view, this integrated optimization problem can be formulated as follows:

$$min\sum_{j}^{N_{agg}}\sum_{g=1}^{N_{bus}}\sum_{i=\min(T_{st})}^{\max(T_{end})}\left[A_j\left(P_{g,i,j}^{ChN}, P_{g,i,j}^{DisN}\right) + D\right] \quad (45)$$

S.t. (38)-(43), (53), (54)

$$P_{g,i}^{DSO} = \sum_{j=1}^{N_{agg}}\left(P_{g,i,j}^{ChN} + P_{g,i,j}^{DisN}\right) \quad (46)$$

It can be seen that (56) is the global constraint for the optimization problem of DSO and aggregators.

### *G. Distributed Algorithm for TE Realization in the Distribution System*

Eq. (56) can be split into two single objective functions using Lagrange Multipliers, $\lambda$, which is also the virtual price signal [26][27]. The Lagrangian function of (55) with coupling constraint (56) can then be formulated in the following.

$$\begin{aligned} &L\left(\boldsymbol{\lambda}, \boldsymbol{P}^{ChN}, \boldsymbol{P}^{DisN}, \boldsymbol{P}^{DSO}\right) = \sum_{j=1}^{N_{agg}} M_j\left(\boldsymbol{P}_j^{ChN}, \boldsymbol{P}_j^{DisN}\right) + \\ &D\left(\boldsymbol{P}^{DSO}\right) + \sum_{g=1}^{N_{bus}}\sum_{i=\min(T_{st})}^{\max(T_{end})}\lambda_{g,i}\left[\sum_{j=1}^{N_{agg}}\left(P_{g,i,j}^{ChN} + P_{g,i,j}^{DisN}\right) - P_{g,i}^{DSO}\right] \end{aligned} \quad (47)$$

In order to solve the above problem, a distributed optimization algorithm [26] is applied to decompose the problem so that the complexity will be greatly reduced. The EVA and DSO will optimize their own schedules but with mutual interests which are linked by the $\lambda$. Then, the EVA's and DSO's optimization problem can be rewritten respectively in the following.

$$min\ A_j\left(P^{ChN}, P^{DisN}\right) + \sum_{g=1}^{N_{bus}}\sum_{i=\min(T_{st})}^{\max(T_{end})}\lambda_{g,i}\left(P_{g,i,j}^{ChN} + P_{g,i,j}^{DisN}\right) \quad (48)$$

S.t. (38)-(43), (53), (54)

$$min\ D\left(P^{DSO}\right) - \sum_{g=1}^{N_{bus}}\sum_{i=\min(T_{st})}^{\max(T_{end})}\lambda_{g,i}P_{g,i}^{DSO} \quad (49)$$

The coordinated control by IMO is realized by updating the $\lambda$. An agreement between EVA and DSO will be reached if $\lambda$ converges. In this work, the sub-gradient method [28] is applied to update $\lambda$ that relies on the multiple iterations of information exchange. The Locational marginal prices for each bus in each iteration can be formulated in the following.

$$\lambda_{g,i}\left(l+1\right)=\lambda_{g,i}\left(l\right)+\beta(l)\left[\sum_{j=1}^{N_{agg}}\left(\boldsymbol{P}_j^{ChN*}+\boldsymbol{P}_j^{DisN*}\right)-P_{g,i}^{DSO*}\right] \quad (50)$$

The convergence condition for the above problem is defined in the following.

$$\left|\lambda_{g,i}\left(l+1\right)-\lambda_{g,i}\left(l\right)\right|\leq\omega \quad (51)$$

In this work, the optimization problem is formulated with YALMIP toolbox under Matlab [29] with the Gurobi solver [30]. In order to solve the problem (53), the Jacobian matrix is derived using MATPOWER [31],.

## IV. Case Study

In this section, the proposed TE method is applied in a representative distribution system of Denmark to help alleviate congestion and voltage violation problems caused by the simultaneous participation of EVs. The parameters of EVs are presented first. Finally, the results and discussions are given.

### A. *Parameter Settings*

In this work, it is assumed that there are two types of EVs managed by two different aggregators in a 10/0.4 kV system which is the same case that specified in [20]. The power transformer capacity allocated to the EVAs is 70 kW. The minimum/maximum voltage $U_{min}/U_{max}$ per bus is assumed to be 0.9 and 1.1 p.u. respectively. The parameters of each type of EV battery are specified in Table II.

TABLE II. EV Battery Parameters [32][33]

| EV Type | $E_b$ | $SOC^{min}$ (%) | $SOC^{max}$ (%) | $P^{MaxCh}$/ $P^{MaxDis}$ | $\eta_{Ch}$ / $\eta_{Dis}$ | $L_c$ | *DoD* |
|---|---|---|---|---|---|---|---|
| 1 | 14 | 20 | 90 | 3.7 | 0.9/ 0.95 | 4000 | 0.8 |
| 2 | 25 | 20 | 85 | 5.28 | 0.9/ 0.95 | 4000 | 0.8 |

The historical/predicted prices for each scenario in the DAM are illustrated in Fig. 4. As the driving pattern of each EV owner can be different, the driving behavior of each EV owner is simulated by two aspects: normal daily driving distance considering driving habit, which results in a different initial SOC and minimum required SOC after departure and various arrival and departure time. The arrival/departure time for different EV is illustrated in Fig. 5.

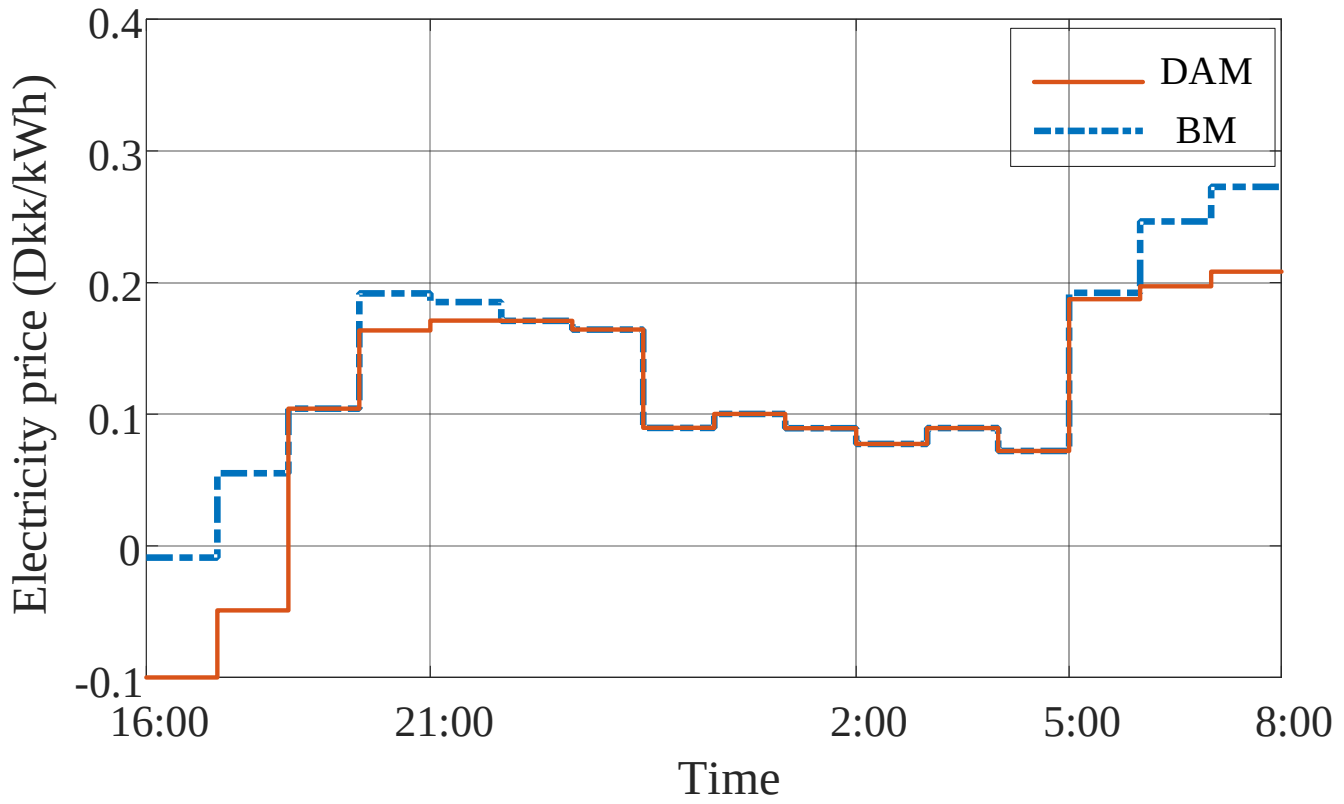

Fig. 4. Historical electricity markets' price.

In Fig. 4, the historical DAM and BM price are compared and is considered as the input for the optimization problem specified in section III. B.

### B. *Aggregator's Optimal Schedule in DAM*

In this part, the deterministic optimization (DO) with the historical prices are presented. In each scenario, the optimization problem was solved by considering BOC (SI) or not (SII). The schedule for each TVVS is illustrated in Fig. 6.

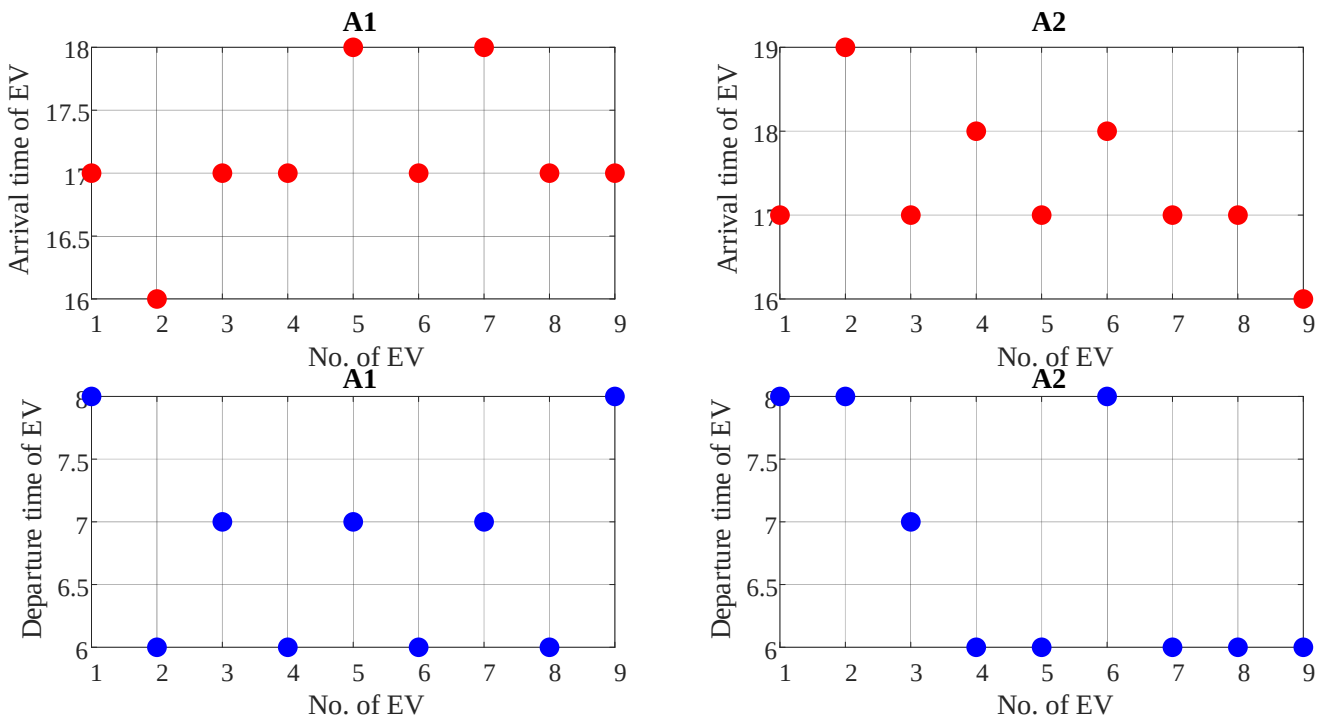

Fig. 5. Arrival/departure time of each EV of the corresponding aggregator.

In Fig. 6, the positive values indicate the charging energy while negative values show the discharging energy. It can be seen that the V2G function will not be used if the BOC is considered.

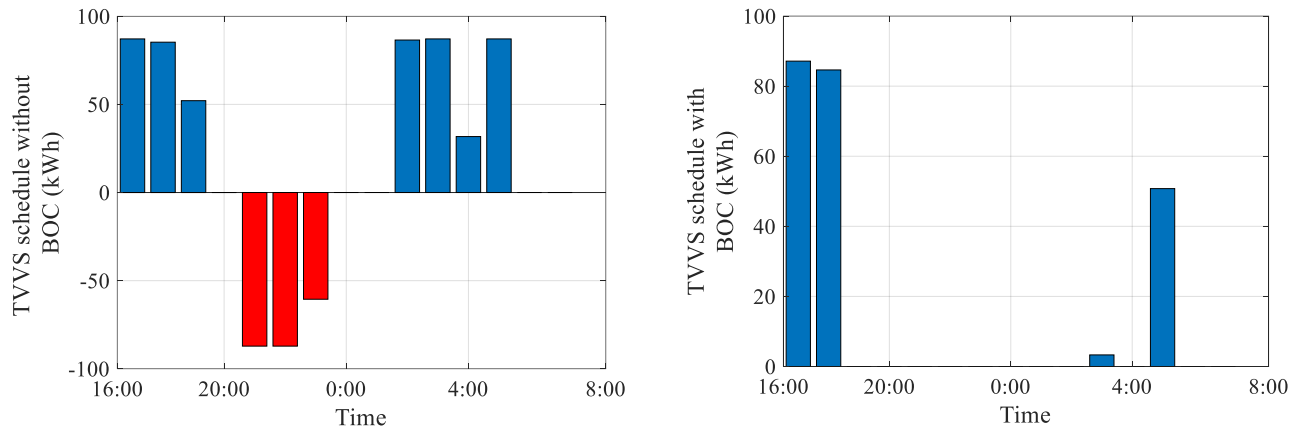

(a) (b)

Fig. 6. Battery optimal schedule. (a) Optimal schedule without considering BOC (Scenario I) (b) Optimal schedule considering BOC (Scenario II).

### C. *Aggregator's Updated Schedule after BM*

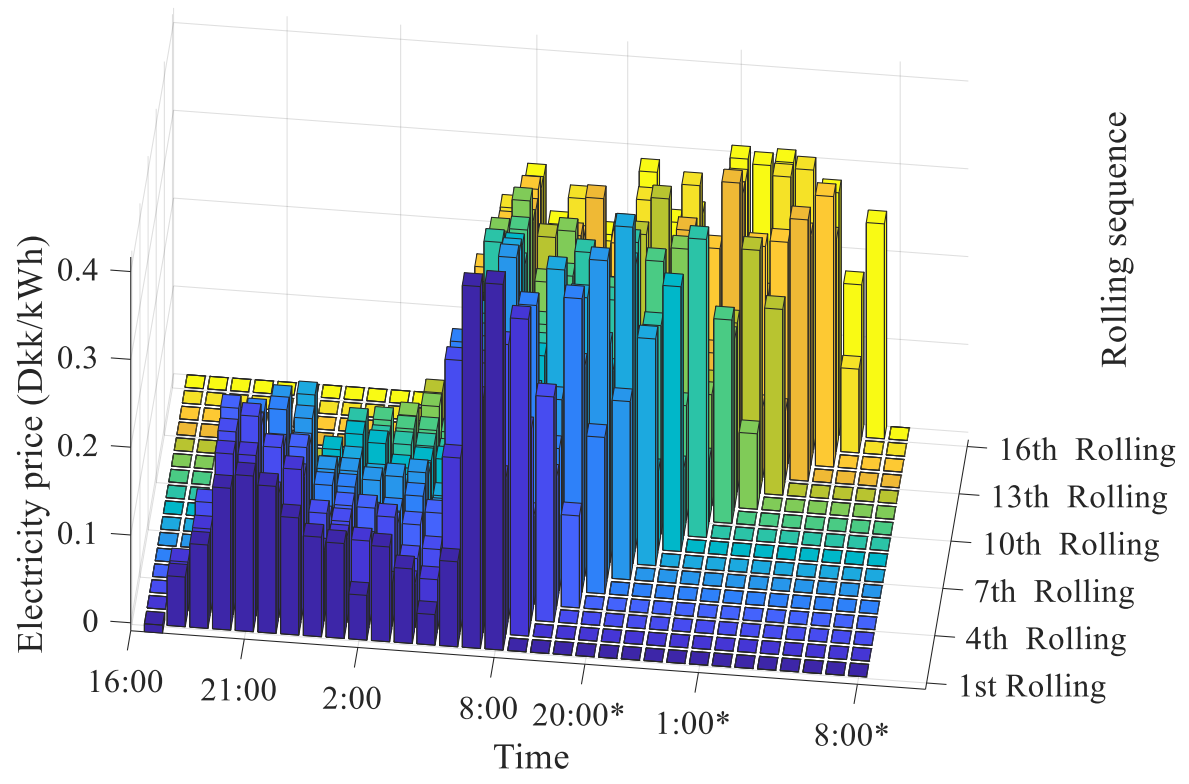

Fig. 7. The predicted BM price for each rolling process.

After bidding in DAM, the EVAs' schedule can also be updated in BM. In this work, it is assumed that the prediction error is increased with a maximum step of 1.5% per hour. In other words, the price deviation is increasing along with the time horizon. The predicted BM price for each rolling procedure is shown in Fig. 7. The optimal schedule before/ after BM and with/without RWO in BM are compared in Fig. 8.

It can be seen in Fig. 7 that the price deviation is increasing along with the time horizon. This is according to the fact that the predicted price will be more accurate if the prediction time is closer to real time.

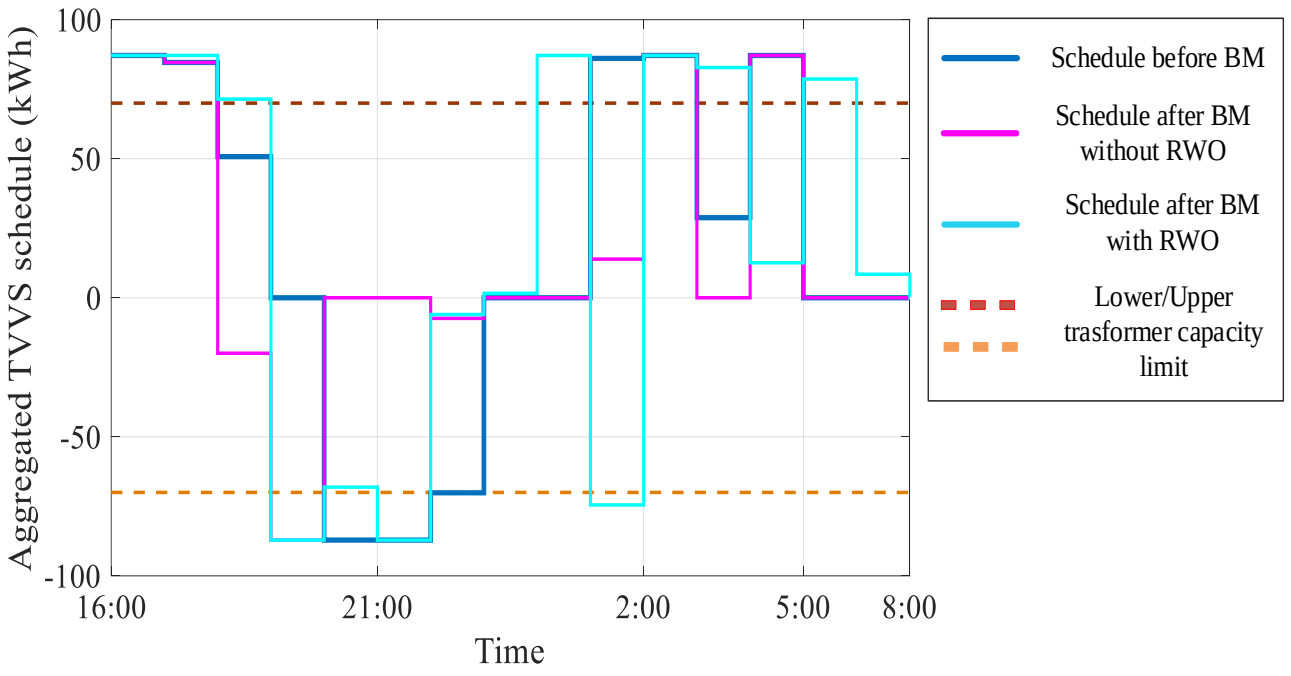

(a)

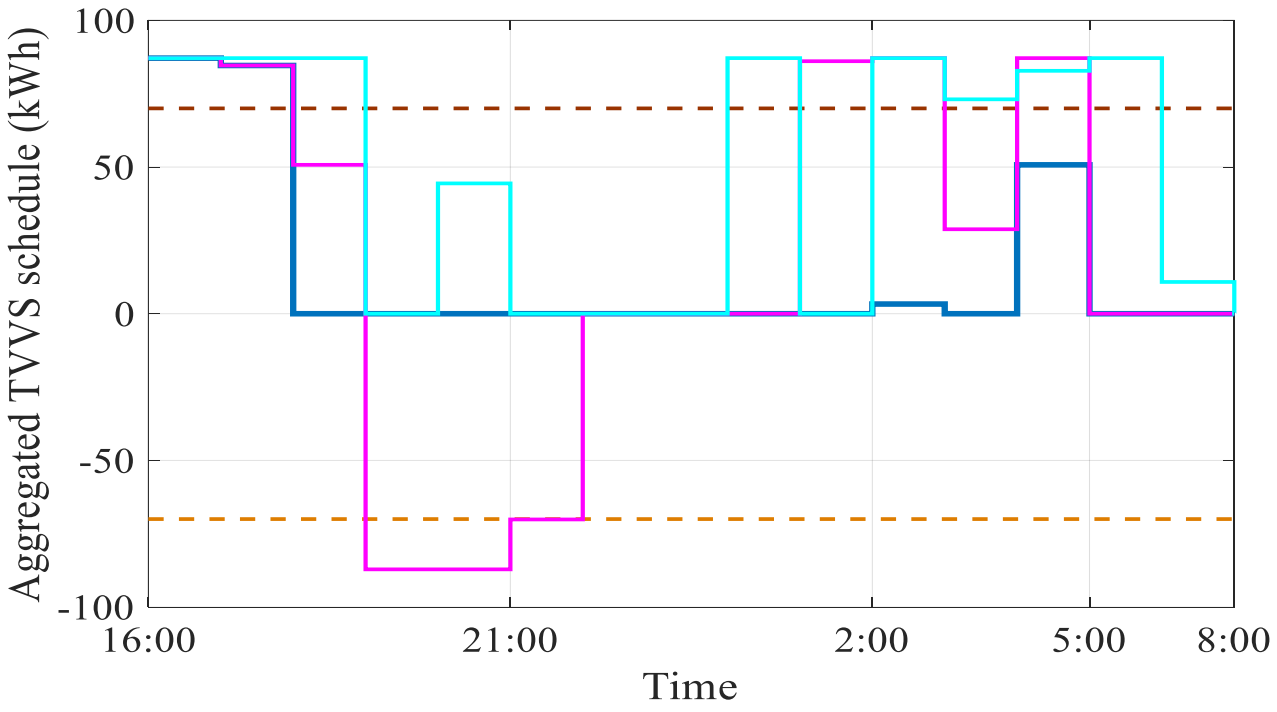

(b)

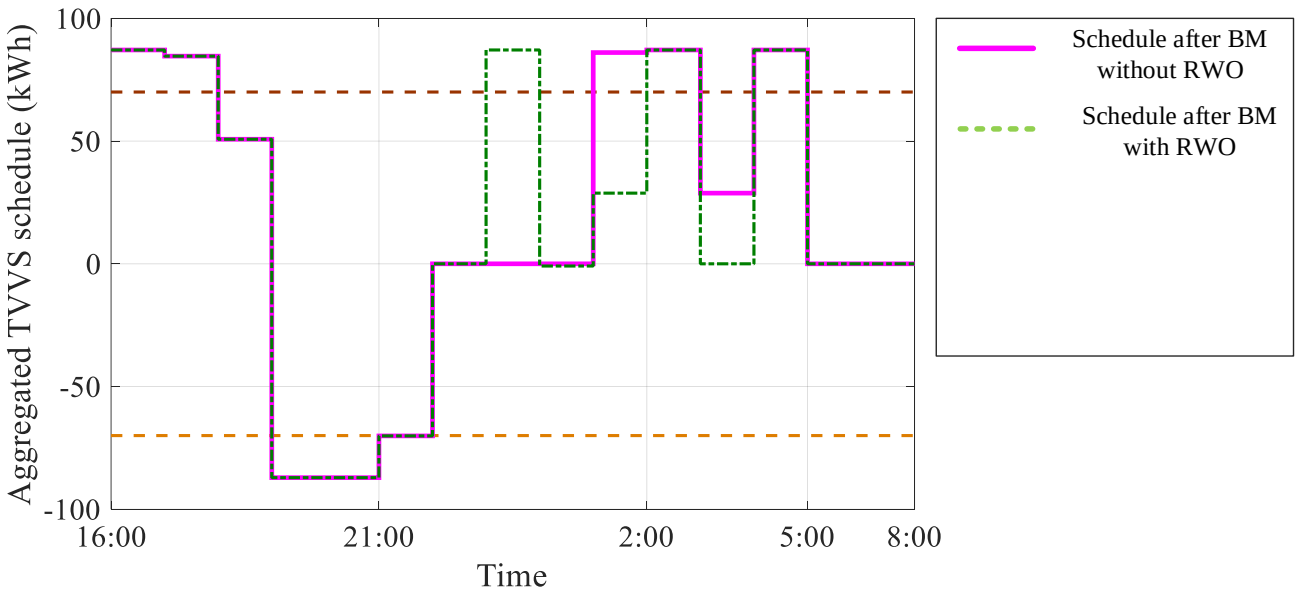

(c)

Fig. 8. The sum of the aggregators' power before and after the BM. (a) Schedule comparison without considering BOC. (b) Schedule comparison considering BOC. (c) Schedule comparison after 1st RWO.

After each rolling process, the first time interval schedule will be selected. The final and aggregated schedule before/after BM without using RWO and using RWO is shown as the blue/pink/cyan line segments in Fig. 9 respectively. Compared Fig. 8 (a) with (b), it can be known that the congestion problem occurs more frequently when the BOC is considered. Hence, the TE should be applied for resolving the problem.

### D. *Modified Schedule after Applying TE*

In this part, the TE is applied for the case with or without considering BOC. If the BOC's impact is neglected, the optimal schedule before and after using TE is shown in Fig. 9. On the contrary, the optimal schedule considering BOC's impact is shown in Fig. 10.

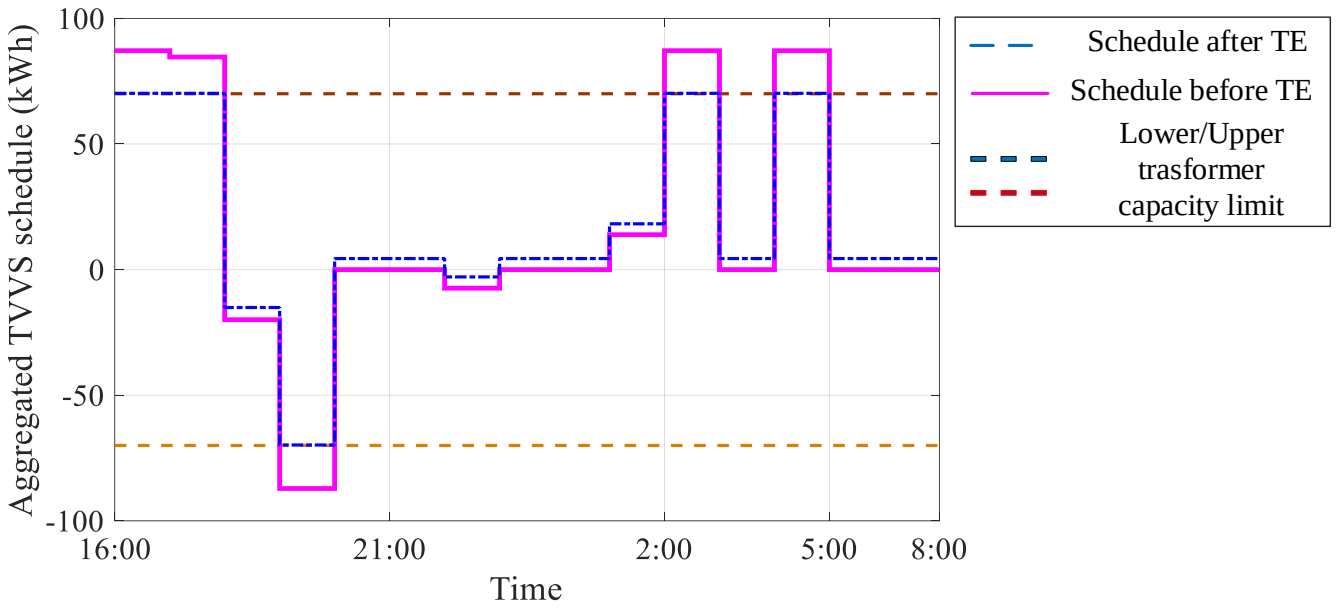

(a)

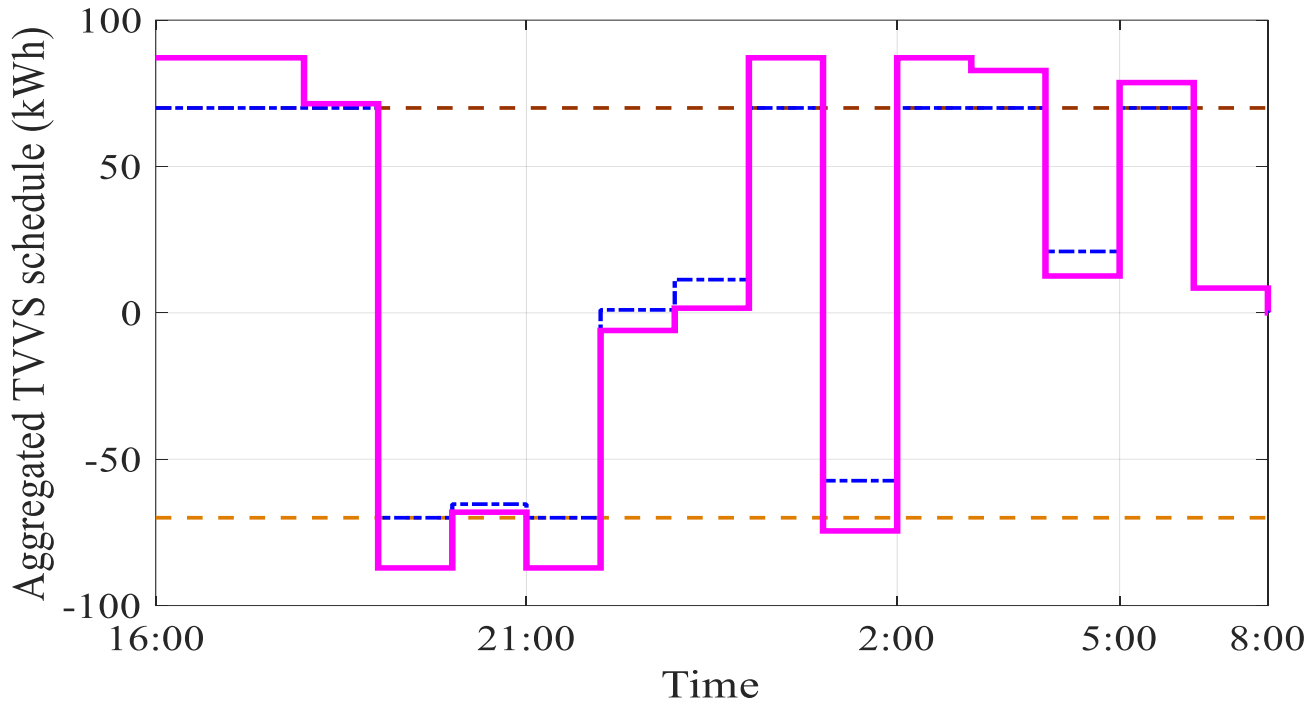

(b)

Fig. 9. The sum of the aggregators' power before and after the TE without considering BOC. (a) Schedule comparison without RWO. (b) Schedule comparison with RWO.

The programme was run on a processor of Intel (R) Core (TM) i7-7700HQ CPU @ 2.80 GHz. The computational time of using TE is 119 seconds. As can be seen in both Fig. 9 and 10, the updated schedule via the TE does not incur any congestion problem.

## V. Conclusion

A continuous operational framework for EVA participation in both DAM and BM is proposed. The BOC is considered as an additional operational cost. To meet the system constraints, the TE method is applied to help congestion management and control voltage violations. To reduce the impact of prediction error in electricity price, a RWO method is adopted to get the optimal schedule for EVA. From the simulation, the V2G solution is still too expensive although the battery cost has declined greatly. The congestion problem in future distribution system would happen more frequently if EV owners are willing to provide down-regulation service, where network constraints must be taken into account during operation. The proposed framework has an advantage in its decentralized structure for implementation and flexibility in handling the operational interests of different parties.

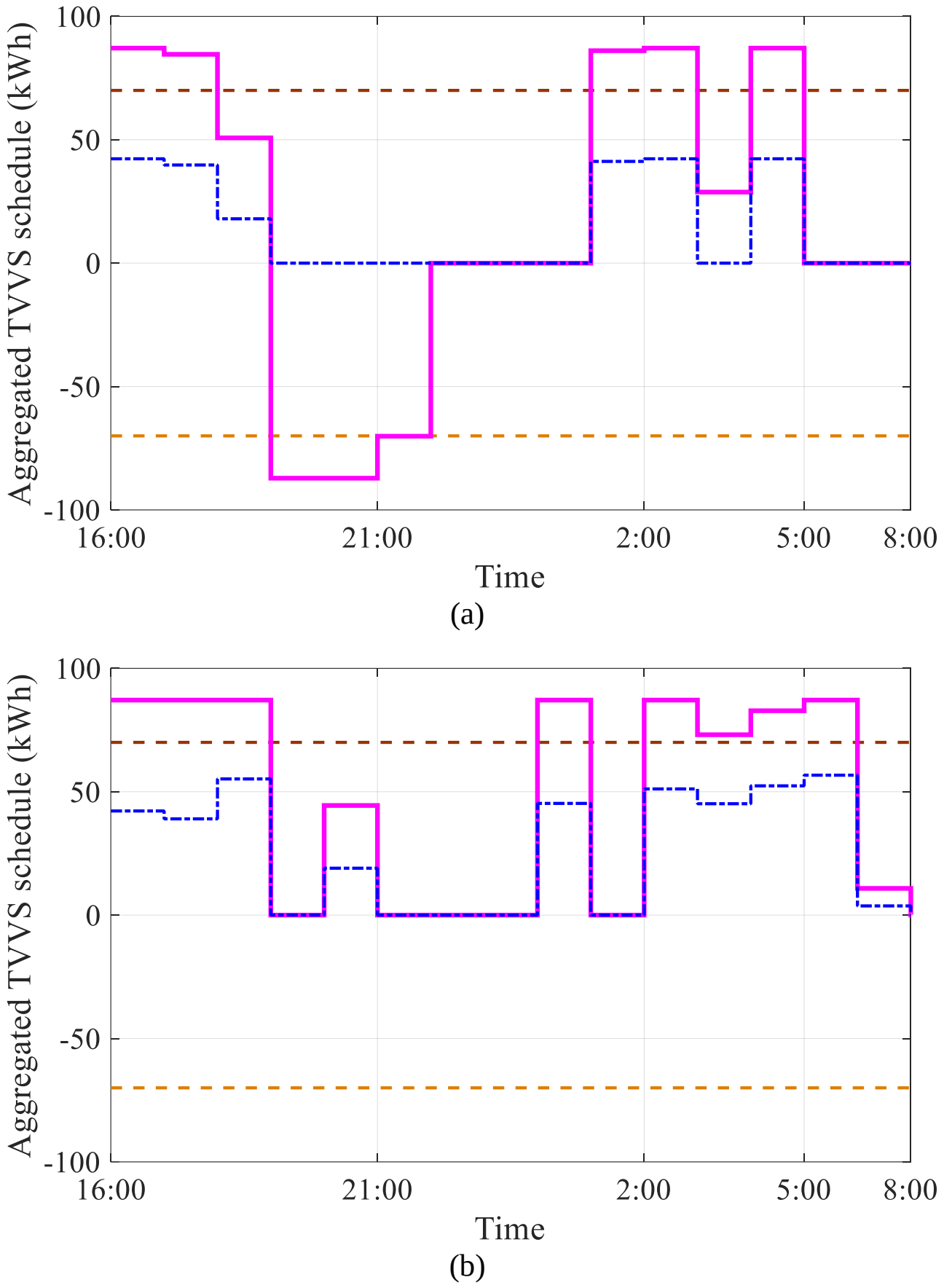

Fig. 10. The sum of the aggregators' power before and after the TE with considering BOC. (a) Schedule comparison without RWO. (b) Schedule comparison with RWO.

## References

[1] United Nations Climate Change. [Online]. Available: https://unfccc.int/process/the-paris-agreement/the-paris-agreement

[2] Y. Parag and B. K. Sovacool, "Electricity market design for the prosumer era," Nat. Energy, vol. 1, no. 4, Mar. 2016.

[3] S. I. Vagropoulos and A. G. Bakirtzis, "Optimal bidding strategy for electric vehicle aggregators in electricity markets," *IEEE Trans. Power Syst.*, vol. 28, no. 4, pp. 4031–4041, Nov. 2013.

[4] C. Jin, J. Tang, P. Ghosh, "Optimizing Electric Vehicle Charging With Energy Storage in the Electricity Market," *IEEE Trans. Smart Grid*, vol. 4, no. 1, pp. 311-320, Mar. 2013.

[5] R. Li, Q. Wu, S. S. Oren, "Distribution Locational Marginal Pricing for Optimal Electric Vehicle Charging Management," *IEEE Trans. Power Syst*, vol. 29, no. 1, pp. 203-211, Jan. 2014.

[6] E. Veldman, R. A. Verzijlbergh, "Distribution Grid Impacts of Smart Electric Vehicle Charging From Different Perspectives," *IEEE Trans. Smart Grid*, vol. 6, no. 1, pp.333-342, Jan. 2015.

[7] N. Y. Soltani, S. J. Kim, G. B. Giannakis, "Real-Time Load Elasticity Tracking and Pricing for Electric Vehicle Charging," *IEEE Trans. Smart Grid*, vol. 6, no. 3, pp. 1303-1313, May 2015.

[8] M. R. Sarker, Y. Dvorkin, and M. A. Ortega-Vazquez, "Optimal participation of an electric vehicle aggregator in day-ahead energy and reserve markets," *IEEE Trans. Power Syst.*, vol. 31, no. 5, pp. 3506–3515, 2016.

[9] Z. Liu, Q. Wu, S. Huang, L. Wang, Mohammad, "Optimal Day-ahead Charging Scheduling of Electric Vehicles through an Aggregative Game Model," *IEEE Trans. Smart Grid*, vol. 9, no. 5, pp. 5173 - 5184 , Sept. 2018.

[10] B. Sun, Z. Huang, X. Tan, D. H. K. Tsang, "Optimal Scheduling for Electric Vehicle Charging With Discrete Charging Levels in Distribution Grid," *IEEE Trans. Smart Grid*, vol. 9, no. 2, pp. 624-634, Mar. 2018.

[11] P. Sánchez-Martín, S. Lumbreras, A. Alberdi-Alén, "Stochastic Programming Applied to EV Charging Points for Energy and Reserve Service Markets," *IEEE Trans. Power Syst.*, vol. 31, no. 1, pp. 198-205, Jan. 2016.

[12] R.-C. Leou, "Optimal Charging/Discharging Control for Electric Vehicles Considering Power System Constraints and Operation Costs," *IEEE Trans. Power Syst.*, vol. 31, no. 3, pp. 1854-1860, May 2016.

[13] Maigha, M. L. Crow, "Electric Vehicle Scheduling Considering Co-optimized Customer and System Objectives," *IEEE Trans. Sustain. Energy*, vol. 9, no. 1, pp. 410-419, Jan. 2018.

[14] J. Hu, G. Yang, H. W. Bindner, Y. Xue, "Application of Network-Constrained Transactive Control to Electric Vehicle Charging for Secure Grid Operation," *IEEE Trans. Sustain. Energy*, vol. 8, no. 2, pp. 505–515, Apr. 2017.

[15] K. Kok and S. Widergren, "A society of devices: Integrating Intelligent Distributed Resources with Transactive Energy," *IEEE Power Energy Mag.*, vol. 14, no. 3, pp. 34–45, May-Jun 2016.

[16] N. Korolko, Z. Sahinoglu, "Robust Optimization of EV Charging Schedules in Unregulated Electricity Markets," *IEEE Trans. Smart Grid*, vol. 8, no. 1, pp. 149-157, Jan. 2017.

[17] H. Ding, Z. Hu, Y. Song, "Rolling Optimization of Wind Farm and Energy Storage System in Electricity Markets," *IEEE Trans. Power Syst.*, vol. 30, no. 5, pp. 2676-2684, Sept. 2015.

[18] S. I. Vagropoulos, G. A. Balaskas, A. G. Bakirtzis, "An Investigation of Plug-In Electric Vehicle Charging Impact on Power Systems Scheduling and Energy Costs," *IEEE Trans. Power Syst.*, vol. 32, no. 3, pp. 1902-1912, May 2017.

[19] W. Gu, Z. Wang, Z. Wu, Z. Luo, Y. Tang, J. Wang, "An Online Optimal Dispatch Schedule for CCHP Microgrids Based on Model Predictive Control," *IEEE Trans. On Smart Grid*, vol. 8, no. 5, pp. 2332-2342, Sept. 2017.

[20] J. Hu, G. Y, C. Ziras, K. Kok, "Aggregator operation in the balancing market through network-constrained transactive energy," *IEEE Trans. on Power System*, vol. 34, no. 5, pp. 4071-4080, Sept 2019.

[21] C. Bang, F. Fock, M. Togeby, "The existing Nordic regulating power market," FlexPower WP1 – Report 1, 2012.

[22] HEP/HEP, "Regulation C2: The balancing market and balance settlement," Energinet.dk, Tonne Kjærsvej 65, DK-7000 Fredericia 2008.

[23] C. A. Floudas, *Nonlinear and mixed-integer optimization: fundamentals and applications*. Oxford University Press, 1995.

[24] D. Mignone, "The really big collection of logic propositions and linear inequalities," Automatic Control Laboratory, Zurich, Switzerland, Tech. Rep. AUT01-11, Feb. 27, 2002.

[25] S. Bolognani and S. Zampieri, "On the existence and linear approximation of the power flow solution in power distribution networks," *IEEE Trans. Power Syst.*, vol. 31, no. 1, pp. 163–172, Jan 2016.

[26] M. Diekerhof, S. Vorkampf, A. Monti, "Distributed Optimization Algorithm for Heat Pump Scheduling," 2014 5th IEEE PES Innovative Smart Grid Technologies Europe (ISGT Europe), October 12-15, Istanbul, 2014.

[27] S. Mhanna, A. C. Chapman and G. Verbič, "A fast distributed algorithm for large-scale demand response aggregation," *IEEE Trans. Smart Grid*, vol. 7, no. 4, Jul. 2016.

[28] S. Boyd and A. Mutapcic, "Subgradient methods," *Lecture notes of EE364b, Stanford University, Winter Quarter*, vol. 2007, 2006.

[29] J. Lofberg, "Yalmip : A toolbox for modeling and optimization in matlab," in In Proceedings of the CACSD Conference, (Taipei, Taiwan), 2004.

[30] "Gurobi optimizer." http://www.gurobi.com/.

[31] R. D. Zimmerman, C. E. Murillo-Sanchez, and R. J. Thomas, "Matpower Steady-state operations, planning, and analysis tools for power systems research and education," *IEEE Trans. Power Syst.*, vol. 26, no. 1, pp. 12–19, Feb. 2011.

[32] C. Curry, "Lithium-ion Battery Costs and Market-Squeezed margins seek technology improvements & new business models," Bloomberg New Energy Finance – BNEF report, July 2017.

[33] Y. Mou, H. Xing, Z. Lin, M. Fu, "Decentralized Optimal Demand-Side Management for PHEV Charging in a Smart Grid," *IEEE Trans. Smart Grid*, vol. 6, no. 2, pp. 726-736, Mar. 2015.